\documentclass[aps,pre,superscriptaddress]{revtex4}

\usepackage{graphicx}
\usepackage[utf8]{inputenc}
\usepackage[english]{babel}
\usepackage{amsmath}
\usepackage{graphicx,enumerate}
\usepackage{epsfig}
\usepackage{subfig}
\usepackage{amssymb}
\begin{document}

\title{Statistical mechanics of systems with long-range interactions and negative absolute temperature}

\author{Fabio Miceli}
\affiliation{Dipartimento di Fisica, Universit\`a di Roma Sapienza, P.le Aldo Moro 5, 00185, Rome, Italy}
\affiliation{Present affiliation: Mathematics Institute, University of Warwick, Coventry CV4 7AL, United Kingdom}

\author{Marco Baldovin}
\affiliation{Dipartimento di Fisica, Universit\`a di Roma Sapienza, P.le Aldo Moro 5, 00185, Rome, Italy}

\author{Angelo Vulpiani}
\affiliation{Dipartimento di Fisica, Universit\`a di Roma Sapienza, P.le Aldo Moro 5, 00185, Rome, Italy}
\affiliation{Centro Linceo Interdisciplinare ``B. Segre'', Accademia dei Lincei, Rome, Italy}

\date{\today}

\begin{abstract}
A Hamiltonian model living in a bounded phase space and with long-range 
interactions is studied. It is shown, by analytical computations, that there 
exists an energy interval in which the microcanonical entropy is a decreasing 
convex function of the total energy, meaning that ensemble equivalence is violated in
a negative-temperature regime. The equilibrium properties of the model 
are then investigated by molecular dynamics simulations: first, the caloric 
curve is reconstructed for the microcanonical ensemble and compared to the 
analytical prediction, and a generalized Maxwell-Boltzmann distribution for the 
momenta is observed; then, the nonequivalence between the microcanonical and 
canonical descriptions is explicitly shown. Moreover, the validity of 
Fluctuation-Dissipation Theorem is verified through a numerical study, also at 
negative temperature and in the region where the two ensembles are 
nonequivalent.
\end{abstract}

\maketitle

\section{Introduction}

It is well known that in physical systems with long-range interactions the 
equivalence of statistical ensembles can fail~\cite{dauxois02, campa09}, meaning that there 
exist equilibrium states described by the microcanonical probability density 
function (p.d.f.) that do not correspond to any state described by the canonical 
one: in other words, the average of a macroscopic observable can give different 
results, at the same temperature $T$, if the system is isolated or closed. This 
phenomenon can be observed in a large variety of physical contexts including 
self-gravitating systems, perfect fluids in two dimensions and spin models with 
mean field 
interactions~\cite{padmanabhan90, lyndenbell68, lyndenbell99, chavanis02, chavanis06,
ramirezhernandez09, hovhannisyan17}.

 For long-range interacting systems, inequivalence of statistical ensembles is 
due to the lack of additivity of the total energy $E$, which can result in a 
change of concavity for the entropy $S(E)$ (i.e. in negative specific heat): 
since the Legendre-Fenchel transform that relates the free energy $F(T)$ to 
$S(E)$ is not invertible in this case, there is no one-to-one correspondence 
between the microcanonical and the canonical 
description~\cite{touchette09,patelli14}. In particular, it can be shown that
the caloric curve $T$ vs $E$ in the canonical ensemble can be obtained from the microcanonical
analogue through the Maxwell construction, i.e. by replacing the free 
energy with its convex envelope. The situation is somehow reminiscent
of that arising in van der Waals equation for non ideal gases, obtained 
by a mean field approach, where the pressure  is  a non-monotonic function of 
the volume: Maxwell construction was indeed introduced in this context,
in order to describe the coexistence of two different phases in the
``unphysical'' region at negative compressibility~\cite{ma85, huang88}.
Let us stress, however, that in this case there is no real ensemble inequivalence:
as discussed in~\cite{lebowitz66}, the negative compressibility is a mere
consequence of the homogeneity assumption, and it does not appear if 
this hypothesis is relaxed; the equation of state can be derived through a rigorous
limit procedure on the interaction potential (the so-called van der Waals scaling),
and no negative-compressibility region is found. Ensemble inequivalence 
arises instead if the long-range interaction part of the potential is mean-field~\cite{kiessling95}.

The aim of this paper is to investigate the statistical features of a system 
with ensemble inequivalence in a negative temperature regime: in particular we
show that the momenta, at equilibrium, follow a generalized Maxwell-Boltzmann
distribution, and that a Fluctuation-Dissipation theorem holds.

Let us recall that a system with Hamiltonian $H(\mathbf{X})$, where $\mathbf{X}$ 
is a point in the phase space $\Omega$, is said to have a negative absolute 
temperature if its Boltzmann entropy 
\begin{equation}
\label{eq:entr}
 S(E)=\ln \int_{\Omega} d\mathbf{X}\,\delta(H(\mathbf{X})-E)\,
\end{equation}
is a decreasing function of the energy $E$ in a certain energy interval (here 
and in the following, we set the Boltzmann constant $k_B$ equal to 1): this 
essentially means that the volume of the accessible phase-space region shrinks 
when increasing the energy of the system~\cite{ramsey56}. It is easy to 
understand that Hamiltonians in which the kinetic energy has the usual form, 
quadratic in the momenta, cannot achieve negative temperature: for the ideal gas 
$S(E)\propto \ln E$, and additional degrees of freedom 
cannot change the sign of $dS(E)/dE$.

The possibility of negative absolute temperature states in Hamiltonian systems 
is well established. An important example is already described in the Onsager's 
work on the statistical hydrodynamics of point vortices~\cite{onsager49}; this 
seminal contribution was the starting point for the 
investigation of the dynamical and statistical properties of hydrodynamics 
systems~\cite{marchioro94}. These models are interesting both for applications (e.g. in 
plasma physics) and for their statistical mechanics properties. Let us note, for 
instance, that an example of a system with ensemble inequivalence at 
negative temperature is already studied in~\cite{smith90}: here the considered system
is a guiding-center model for a plasma in a cylindric domain, equivalent to a 2-D point
vortices system; it is shown that the equilibrium solution of the mean-field Vlasov
equation, in a high-energy range at negative temperature, presents
ensemble inequivalence. Rigorous results on the statistical mechanics of point vortices
in bounded domains have been proven~\cite{kiessling93,caglioti92,caglioti95}, relating
the inequivalence of statistical ensembles to the non-uniqueness of the solution,
at high energy and for some kind of bounded domains in two dimensions,
of the mean-field equation describing the system. Recently, the problem of equivalence
of statistical ensembles for systems of point voritices on a spherical surface has been
also addressed~\cite{kiessling12}.

The relevance of the presence of negative temperature has been investigated in another important Hamiltonian 
model, the Discretized Nonlinear Schr\"{o}dinger Equation: their emergence in 
out-of-equilibrium conditions have been recently investigated~\cite{iubini12, 
iubini13,iubini17}. Experimental evidence of negative temperature equilibrium 
states, on the other hand, dates back to the studies on nuclear spins by 
Purcell, Pound and Ramsey in the 50's~\cite{purcell51, ramsey56}. In more recent 
times, experiments have shown that negative  temperature is also present in 
systems of cold atoms~\cite{braun13}.

The structure of this paper is the following: in Section~\ref{sec:model} we 
describe the Hamiltonian model we are interested in, and define the 
mechanical observable that we will use to measure the temperature of the system 
in numerical simulations; in Section~\ref{sec:equil} we present our results at equilibrium, showing that it is possible 
to have ensemble inequivalence at negative temperature; Section~\ref{sec:fdr} is 
devoted to the study of response at negative temperature (in particular, the 
validity of the Fluctuation-Dissipation Theorem is verified); in 
Section~\ref{sec:conc} we briefly sketch our conclusions. 
Appendix~\ref{sec:appld} contains the calculations that lead to 
our analytical results, while Appendix~\ref{sec:algo} illustrate the details 
of the algorithm we have used for our simulations.

\section{The model}
\label{sec:model}

Let us consider a Hamiltonian system consisting of $N$ degrees of freedom described by conjugated 
coordinates $\{p_i, \theta_i\}$, $i=1,...,N$. Positions $\{\theta_i\}$ and 
momenta $\{p_i\}$ are both angular variables, meaning that they are constrained 
in the interval $(-\pi,\pi]$ with periodic boundary conditions.
The total Hamiltonian reads:

\begin{subequations}
\begin{alignat}{4}
 \label{eq:ham}
 H(\mathbf{X})&=\sum_{i=1}^N \left(1 - \cos{p_i} \right) - Nv(m)\\
  v(m)&=\frac{J}{2}m^2 + \frac{K}{4}m^4 + \text{const.}\label{eq:poten}
\end{alignat}
\end{subequations}
where $m$ is the modulus of the magnetization vector defined as
\begin{equation}
\label{eq:m}
 \mathbf{m}(\{\theta_i\})\equiv \frac{1}{N}\left(\sum_{i=1}^N \cos{\theta_i}\,, \sum_{i=1}^N \sin{\theta_i} \right)\,.
\end{equation}
$J$ and $K$ are parameters that can assume, in general, both positive and 
negative values; the additive constant in Eqn.~\eqref{eq:poten} is actually 
unessential for the dynamics: in the following we will choose it in such a way 
that the minimal energy achievable by the system is zero.

First, let us comment on the unusual dependence of Hamiltonian~\eqref{eq:ham} on 
the momenta. As discussed in the Introduction, negative temperature equilibrium 
states cannot be observed in mechanical systems ruled by quadratic kinetic 
energy; a more promising class of models is obtained by considering ``modified'' 
kinetic terms, defined as periodic functions of the momenta: in this way the 
phase-space is bounded and we can expect the presence of an energy range with
decreasing entropy. This choice has been considered in previous 
works about negative temperature \cite{cerino15,baldovin17,baldovin18}, and its 
consistence has been checked in different contexts. Let us just notice here that 
in the small energy limit such terms reduce to the usual quadratic ones.

The potential term defined by Eqn.~\eqref{eq:poten} is the one that characterize 
the so-called ``Generalized Hamiltonian Mean Field'' (GHMF) model 
\cite{debuyl05} describing the mean field interactions between magnetized 
rotators. The GHMF is an extended version of the Hamiltonian Mean Field 
model~\cite{antoni95} that includes also a quartic dependence on the 
magnetization; among other interesting properties, this system is paradigmatic 
for the study of inequivalence between canonical ad microcanonical ensembles and 
has been extensively studied in past years \cite{bouchet08, campa09, 
baldovin18_lr}.

Let us note that the bound phase space (due to the specific shape of
the ``kinetic'' terms) and the long-range interactions are two features that our model shares with 
two dimensional vortex systems in bounded domains~\cite{onsager49, smith90, marchioro94}.
Therefore it is quite natural to wonder about the possibility of ensemble inequivalence in the negative temperature regime.

Hamiltonian $H$ depends on the angular positions through the magnetization 
modulus $m$, which is maximal ($m=1$) when all the ``modified'' rotators are 
parallel and vanishes if their angular positions are homogeneously distributed in 
$(-\pi,\pi]$; the $N$ factor in front of the potential is needed in order to 
ensure the extensivity of the system (Kac's prescription). Let us stress that
for the system~\eqref{eq:ham} it is possible to write down an analytical computation of the equilibrium 
properties using large deviation approaches \cite{patelli14}: the procedure is quite 
similar to the one presented in \cite{campa09}, and it is explicitly carried out in 
Appendix \ref{sec:appld}.

In order to compare the theoretical prediction for the equilibrium states of the 
system to the results of numerical simulations, as a preliminary step we need to 
define a proper mechanical observable for the inverse temperature~$\beta$. In 
Hamiltonian systems with the usual kinetic energy $\sum_i p_i^2/2m$ one would 
exploit the equipartition theorem \cite{ma85,huang88} and measure $T=1/\beta$ as the 
(temporal) average $\langle p_i^2/m \rangle$ -- assuming that ergodic hypothesis 
holds. Since model \eqref{eq:ham} is not quadratic in the momenta we cannot 
follow this approach.

The microcanonical p.d.f. for the momentum of the $j$-th particle is given by
\begin{equation}
 \label{eq:rho}
 \rho_j(p|E)=\frac{1}{\omega(E)}\int_{\Omega} d\mathbf{X}\, \delta(H(\mathbf{X})-E)\delta(p-p_j)
\end{equation}
where $\mathbf{X}$ is a point in the phase space $\Omega$ and $\omega(E)$ 
represents the density of states (d.o.s.)
\begin{equation}
 \omega(E)=\int_{\Omega} d\mathbf{X}\, \delta(H(\mathbf{X})-E)\,.
\end{equation}
P.d.f. \eqref{eq:rho} can be evaluated as 
discussed in \cite{cerino15}: let us split the phase space into the product 
$\Omega=\Gamma_j\times \Omega'$, where $\Gamma_j\equiv(-\pi,\pi]$ is the interval 
to whom variable $p_j$ belongs; we get 
\begin{equation}
\label{eq:rho_1}
\begin{aligned}
 \rho_j(p|E)&=\frac{1}{\omega(E)}\int_{\Omega'}d\mathbf{X}'\,\delta\left(H'(\mathbf{X'})+k(p_j)-E\right)\\
 &=\omega'(E-k(p))/\omega(E)\\
 &=\exp\left[S'(E-k(p))-S(E)\right]\\
\end{aligned}
\end{equation}
where $k(p)=1-\cos(p)$ is the ``kinetic'' contribution of the single particle,
so that $H'(\mathbf{X}')=H(\mathbf{X})-k(p_j)$ does not depend on $p_j$. 
Here $\omega'(E)$ and $S'(E)$ are the d.o.s. and the microcanonical entropy for the 
Hamiltonian $H'(\mathbf{X}')$, and we have used the definition of entropy~\eqref{eq:entr}.
Expanding equation~\eqref{eq:rho_1} in $k(p)$ 
($E$ is of the order of $N$ and $k(p)$ is a bounded function, so that $|k(p)|\ll 
E$) we finally get
\begin{equation}
\label{eq:equil}
 \rho_j(p|E)\propto \exp\left[-\beta k(p)\right]
\end{equation}
(assuming that $dS'/dE = dS/dE \equiv \beta$ in the thermodynamical limit).
In order to measure the temperature of the system we can study the observable
\begin{equation}
\label{eq:obs}
 \langle \cos p_j \rangle = -I_1(\beta)/I_0(\beta)
\end{equation}
 where  $I_n(x)$ is the $n-$th modified Bessel function of the first kind. It 
can be easily proved that $B(x)=I_1(x)/I_0(x)$ is an odd function, thus it can 
be inverted in order to find $\beta$.

Let us also stress that, since the r.h.s. of the above equation can be positive, 
$\beta$ can assume, in principle, also negative values.

\section{Equilibrium properties}
\label{sec:equil}

Molecular dynamics simulations have been performed in order to study the functional
dependence of the inverse temperature $\beta$ on the energy $E$ of the system. We
used a second-order symplectic Velocity Verlet-like integrator (see Appendix \ref{sec:algo}, case $D=0$), choosing the integration step $\Delta t$
so as to observe total-energy relative fluctuations $\Delta E/E \simeq O(10^{-5})$.
Since we are interested in equilibrium properties, the total integration 
time~$\mathcal{T}$ has always been chosen to be much longer than the typical
characteristic times of the system (which can be shown to be $O(10)$: see e.g.
the plots in Section~\ref{sec:fdr}). 
In order to fix initial conditions with the desired value of the total energy,
some care has to be devoted to the fact that our ``kinetic'' terms are bounded;
first, one has to choose the angular positions in such a way that $0< E-Nv(m)< 2N$,
e.g. by repeated uniform extractions over a suitable interval; once the above constraint
is satisfied, momenta can be randomly extracted in such a way that the sum of
the kinetic terms amounts to the residual energy. We also include a small
external potential $v_{ext}(\mathbf{X})=\cos(\theta_1)$, acting only on the
first particle, in order to break the angular symmetry of the magnetization
$\mathbf{m}$. All interesting observables are computed as temporal averages:
\begin{equation}
\langle A(\mathbf{X}) \rangle \simeq \frac{1}{\mathcal{T}}\int_0^{\mathcal{T}}A(\mathbf{X}(t))dt 
\end{equation}

\begin{figure}
 \centering
 \includegraphics[width=0.5\textwidth]{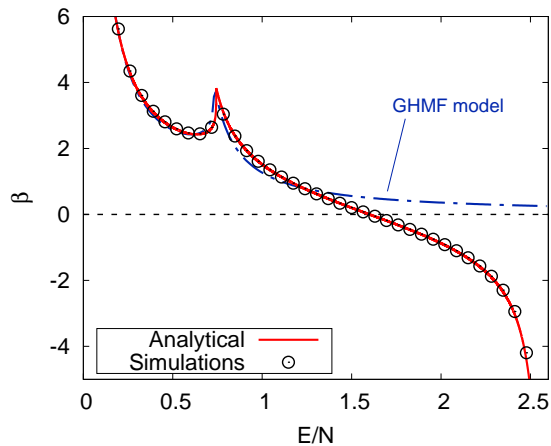}
 \caption{(Color online) Equilibrium caloric curve $\beta(E/N)$ for 
model~\eqref{eq:ham} with $J=0.5$, $K=1.4$. Analytical prediction (red solid 
line) and results of molecular dynamics simulations (black circles) are shown. 
Parameters: $N=1000$, $\Delta t=0.25$, $\mathcal{T}=10^8$. The caloric curve of the GHMF model 
with the same parameters (dash-dotted blue curve) is shown for comparison.} 
\label{fig:ferro}
\end{figure}

In Fig.~\ref{fig:ferro} we show the theoretical caloric curve $\beta(E/N)$ 
for a choice of $J$ and $K$ that leads to inequivalence of statistical ensembles
(namely, to the existence of microcanonical equilibrium states with negative
specific heat that have no canonical counterpart). In this case both $J$
and $K$ are positive, so that the interacting potential has
a minimum for $m=1$ (``ferromagnetic'' limit).

Let us notice that the values of $\beta$ measured in numerical simulations through the observable 
\eqref{eq:obs} show a very good agreement with the analytical prediction.

The figure also shows the difference between the system we are considering
and the corresponding GHMF model with the same parameters. At low energies the behavior of the 
two systems is quite similar, due to the fact that the kinetic terms are equal up
to order $O(p_i^3)$ when energies are small. The scenario changes for large values 
of $E/N$: in this case, the ``bounded phase-space'' version of the GHMF 
model can reach equilibrium states with $\beta \equiv dS/dE <0$, while the 
original GHMF model behaves like an ideal gas whose (positive) $\beta$ 
asymptotically tends to zero (infinite temperature).

\begin{figure}
 \includegraphics[width=0.5\textwidth]{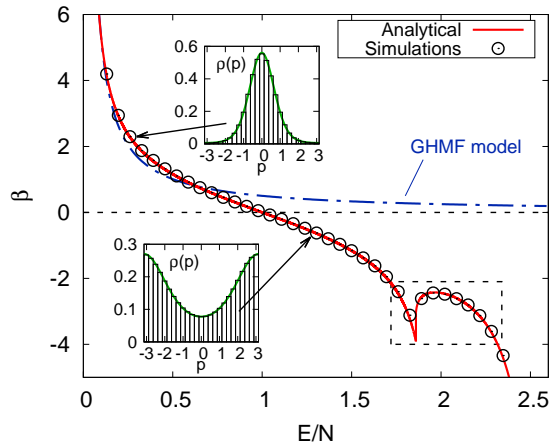}
 \caption{(Color online) Equilibrium caloric curve for model~\eqref{eq:ham} with 
the same parameters of Fig.~\ref{fig:ferro}, but $J=-0.5$, $K=-1.4$. Insets: 
single-particle momentum distributions corresponding to the two cases pointed by 
the arrows; green solid lines represent best fits for the equilibrium 
p.d.f. \eqref{eq:equil}. The fragment of the curve in the dashed rectangle 
is the same shown in Fig.~\ref{fig:zoom}} \label{fig:antif}
\end{figure}

Fig.~\ref{fig:antif} shows a different example, in which both $J$ and $K$ are 
chosen to be negative; in this case $v(m)$ is minimum at $m=0$
(``antiferromagnetic'' limit).

As can be seen in the figure, in a certain energy range 
negative microcanonical specific heat and negative values of $\beta$ coexist -- 
meaning that microcanonical and canonical ensembles are inequivalent for some negative 
temperatures. The microcanonical specific heat of the corresponding GHMF model 
is, instead, always positive, and the statistical ensembles are always equivalent.
In the insets of Fig.~\ref{fig:antif} we also show the single-particle momentum distribution
in a couple of cases (one at positive and one at negative $\beta$): the empirical distribution 
and the analytical expression for the p.d.f. given by Eqn.~\eqref{eq:equil} are in excellent agreement.

\begin{figure}
 \includegraphics[width=0.5\textwidth]{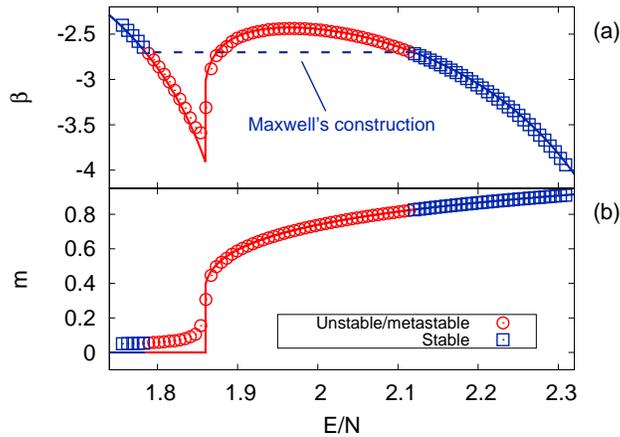}
 \caption{(Color online) (a) Fragment of the caloric curve shown in 
Fig.~\ref{fig:antif}. Blue squares represent stable equilibrium states for which 
ensemble equivalence holds; red circles stand for unstable or metastable 
microcanonical states with no canonical counterpart. The dashed blue line is the 
Maxwell's construction. 
(b): the modulus $m$ of the magnetization of the system is shown in the same conditions of panel (a). 
Solid lines represent, in both panels, the theoretical
prediction.} \label{fig:zoom}
\end{figure}

Let us now focus on the energy range in which statistical ensembles are inequivalent in this case.
Fig.~\ref{fig:zoom} shows the detail of this region. In panel (a), the microcanonical
caloric curve is compared to the results of numerical simulations. The dashed line is the
Maxwell's construction, which individuates the transition temperature $1/\beta^*$
in the canonical ensemble \cite{touchette09}:
all the microcanonical equilibrium states with energies corresponding to the Maxwell's
construction (red circles in the figure) are metastable or unstable \cite{campa09} and have no
equivalent counterpart in the canonical ensemble.
The microcanonical caloric curve also shows a first order phase transition, which 
can be understood by looking at panel (b): the systems goes from a regime with 
$m=0$ to a magnetized phase with $m>0$.

Ensemble inequivalence can be checked through numerical simulations. We start 
from an initial condition whose energy corresponds to an unstable microcanonical 
state, and we let the system evolve through a molecular dynamics simulation for 
a long time $\mathcal{T}_0$. So far the system is isolated, and we can measure 
the inverse temperature $\beta$ of the corresponding microcanonical state by 
studying the above discussed observable. The final configuration is then taken 
as the initial condition for two different evolution protocols: on the one hand 
we let the system evolve further trough a symplectic deterministic integrator, 
as before, for an additional time $\mathcal{T}_1$; on the other hand we can 
perform a stochastic evolution of the system, for the same time, in order to 
mimic the presence of a thermal reservoir and let the system reach a canonical 
equilibrium state, in which $\beta$ is fixed to the value measured before. The 
details about the integrator are discussed in Appendix \ref{sec:algo}. We 
fix the diffusivity constant $D$ that appears in the generalized Langevin 
Equation \eqref{eq:lang} in such a way that $1/|\beta|D$, the typical time-scale 
of the stochastic dynamics, is comparable to the characteristic times of the 
system.

\begin{figure}
 \includegraphics[width=0.5\textwidth]{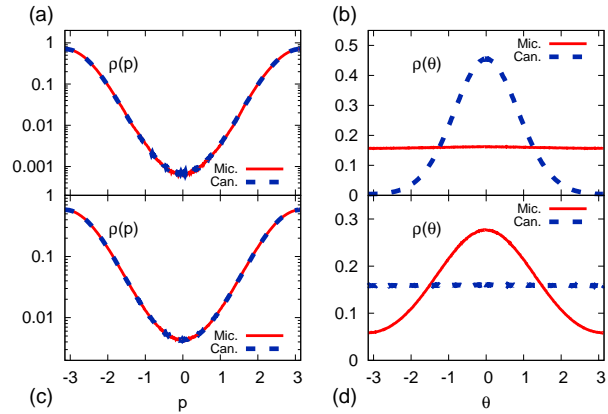}
 \caption{(Color online) Showing ensemble inequivalence in the antiferromagnetic case $J=-0.5$, $K=-1.4$. (a) Distribution of the 
single-particle momentum in the microcanonical ensemble (solid red curve) and 
the corresponding canonical one (dashed blue curve) for $E/N=1.84$. (b) Distribution of the single-particle position, in the same conditions of panel (a). (c-d) Same as (a-b), but for $E/N=1.94$.
Parameters: $N=1000$, $\mathcal{T}_0=10^7$, 
$\mathcal{T}_1=2.5\cdot10^6$, $D=0.02$.} \label{fig:nonequiv}
\end{figure}

The results of our simulations are shown in Fig.\ref{fig:nonequiv} for two 
choices of the initial energy that lead to unstable microcanonical states, 
namely $E/N=1.84$ (panels (a) and (b)) and $E/N=1.94$ (panels (c) and (d)). In 
the left part of the figure, the single-particle momentum distributions, 
microcanonical and canonical, are displayed: in both cases the plots are almost 
identical, meaning that our simulation protocol is able to mimic a canonical 
ensemble with the same temperature of the starting isolated state. On the right 
panels we show the corresponding distributions for the angular positions: in 
both cases we can see a clear difference between the magnetization of the 
original microcanonical state and that of the canonical state at the same 
temperature.

\section{Response theory}
\label{sec:fdr}

In this section we study the response of the system to a small perturbation. In particular we want to check the validity of the well-known Fluctuation-Dissipation Relation (FDR) (see Ref.\cite{marconi08}) that describes the temporal evolution of a mechanical observable, on average, after the system has been perturbed. 

Suppose that at time $t=0$ the microscopic state of the system $\mathbf{X}(0)$ is instantaneously changed into a new configuration $\mathbf{X}'(0)$ by modifying the canonical coordinate (position or momentum) $X_j$:
\begin{equation}
 X_j \rightarrow X'_j=X_j+\delta X_j
\end{equation}
where $|\delta X_j|$ is small on the typical scales of the system. We can follow the evolution of both systems and measure the mechanical observable $A(\mathbf{X})$ in both cases. Defining
\begin{equation}
 \delta A(t)=A(\mathbf{X}'(t))-A(\mathbf{X}(t))\,,
\end{equation}
the FDR ensures that
\begin{equation}
\label{eq:FDR}
\langle \delta A \rangle(t)=-\sum_j\left\langle A(\mathbf{X}(t))\frac{\partial\ln\rho(\mathbf{X})}{\partial X_j}\bigg|_{\substack{t=0}}\right\rangle\delta X_j(0)
\end{equation}
where $\rho(\mathbf{X})$ is the equilibrium p.d.f. and $\langle \cdot \rangle$ represents the corresponding average.

In our case, if we choose $A(\mathbf{X})=\sin p_j$, Eqn.~\eqref{eq:FDR} can be rewritten as:
\begin{equation}
\label{eq:FDR-sinp}
\frac{\langle \delta A\rangle (t)}{\delta p_j(0)} = \beta\langle\sin (p_j(t))\cdot\sin (p_j(0))\rangle
\end{equation}
where a perturbation on the $j$-th momentum has been considered.

The above relation can be simply checked by numerical simulations.
The strategy consists in preparing the system in a certain configuration $\mathbf{X}$, perturbing the momentum of a randomly chosen particle by a quantity $\delta p(0)$ in order to obtain a new configuration $\mathbf{X}'$ and integrating the dynamics of both systems for a time $\tau$. During the evolution one can measure $A(\mathbf{X})$ and $A(\mathbf{X}')$. The procedure is then repeated $M\gg1$ times, starting from the final configuration of the previous evolution of $\mathbf{X}$: since the integrator is symplectic, in this way we are sure to consider initial conditions at the same energy, and we can average over all the measured evolutions.

\begin{figure}
 \includegraphics[width=0.48\textwidth]{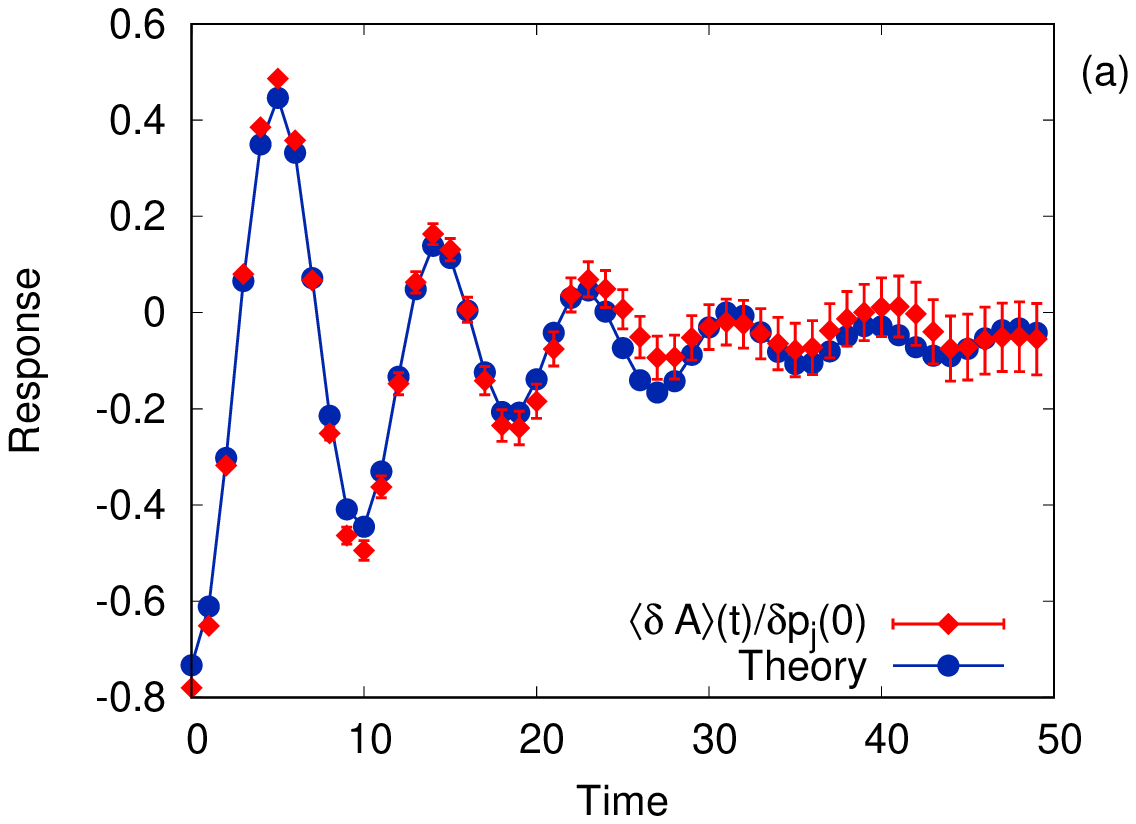}
 \includegraphics[width=0.48\textwidth]{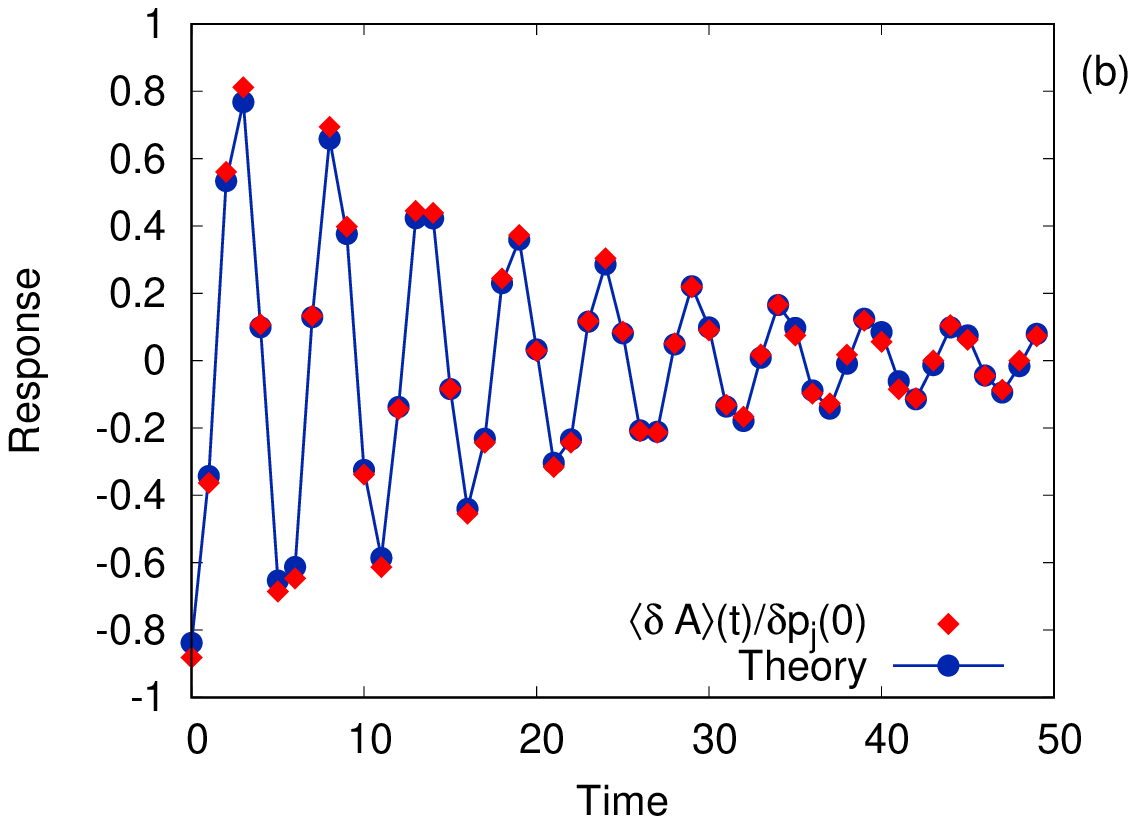}
 \caption{(Color online) Check of the FDR by a direct measure of the l.h.s. of Eqn.~\eqref{eq:FDR-sinp} (red diamonds) and comparison with the theory (r.h.s. of the same equation, blue circles). Panel (a) shows the results for $E/N=1.9$, panel (b) for $E/N=2.3$. Error bars have not been reported in panel (b) since they are invisible on the scale of the plot. Parameters: $J=-0.5$, $K=-1.4$ $N=250$, $M=10^5$, $\tau=50$, $\delta p_i(0)=0.01$.}
 \label{fig:fdr}
\end{figure}

In Fig.~\ref{fig:fdr} we show the results of our numerical simulations for two different choices of $E$. The agreement between the measured values of $\delta A (t)$ and Eqn.~\eqref{eq:FDR-sinp} is clear.

Let us notice that the FDR can be used, in general, to give an operative definition of the inverse temperature $\beta$ of the system through the measure of responses and autocorrelation functions \cite{marconi08}: the presence of negative temperature and ensemble inequivalence does not hinder this possibility.

\section{Conclusions}
\label{sec:conc}
In this paper we studied a Hamiltonian system with long-range interactions and 
modified "kinetic" terms (i.e., the usual $p^2/2$ has been replaced by $1- \cos 
p$ for all particles). In such a model  analytical computations and numerical simulations show the 
existence of an energy range with absolute negative temperature. Comparing the 
results  obtained with microcanonical and canonical ensembles we show  that such 
ensembles are non-equivalent. Even in presence of such non standard features, 
the validity of Fluctuation-Dissipation Theorem is verified through a numerical 
study of the response of the system to a small perturbation.

In spite of the fact that there has been a certain confusion in the literature 
\cite{romero13, dunkel14, hilbert14}, systems with negative temperature  do 
not show any kind of peculiarity and there is no evidence of behaviors in   
disagreement with the basic features established in statical mechanics; this has 
been clearly discussed in several recent papers~\cite{frenkel15, buonsante16, 
puglisi17, swendsen18}. We  briefly list a series of facts showing that systems 
with negative absolute temperature do not show any anomaly~\cite{puglisi17}:
\begin{enumerate}[a)]
 \item the statistical properties of the energy fluctuations are ruled by the inverse
absolute temperature $\beta$ defined by the microcanonical entropy;
\item the p.d.f. of the momentum is given by a generalized Maxwell-Boltzmann distribution
 where $\beta$ appears;
 \item $\beta$  is a measurable quantity, without the appearance of any inconsistency;
 \item in presence of slow variables (e.g. a ``heavy intruder'' in a chain)
it is possible to introduce, in a consistent way,  a Langevin equation~\cite{baldovin18};
 \item possibility of non equivalence of the statistical ensembles~\cite{smith90,caglioti95};
 \item validity of the Fluctuation-Dissipation Theorem.
\end{enumerate}

\acknowledgments{We thank A. Campa for helpful discussions and useful comments.}

\appendix

\section{Large deviation analysis}
\label{sec:appld}

In this Appendix we sketch the large-deviation strategy used to 
determine the equilibrium properties of model~\eqref{eq:ham}. Our 
argument follows quite closely the usual derivation that can be done for the 
GHMF model (see, e.g. \cite{campa09}), with an ad-hoc handling of the ``kinetic'' terms. Let 
us just recall that large deviation approaches can be exploited to study the 
thermodynamic behavior of a system composed of $N$ particles if $n\ll N$ mean 
quantities $\mu_j(\mathbf{X})=\sum_{i=1}^Ng_j(q_i,p_i)/N$, $j=1,...,n$, of the 
phase-space exist such that, for a suitable choice of $\bar{H}$, one has
\begin{equation}
 H(\mathbf{X})=\bar{H}(\mu_1(\mathbf{X}), ..., \mu_n(\mathbf{X}))\,.
\end{equation}
In such a case the following relation holds:
\begin{equation}
\label{eq:entro}
 S(E)=\ln\int \prod_j d\bar{\mu}_j \delta(\bar{H}(\bar{\mu}_1,...,\bar{\mu}_n)-E)\exp[N \bar{s}(\bar{\mu}_1,...,\bar{\mu}_n)]
\end{equation}
where $\bar{s}$ is the so-called ``entropy of macrostates'':
\begin{equation}
 \bar{s}(\bar{\mu}_1,...,\bar{\mu}_n)=\frac{1}{N}\ln\int_{\Omega} d\mathbf{X} \delta(\mu_j(\mathbf{X})-\bar{\mu}_j)\,.
\end{equation}
Furthermore, large deviation theory ensures that the entropy of macrostates can be computed, in the $N\gg1$ limit, as 
\begin{equation}
 \bar{s}=\inf_{\lambda_1, ..., \lambda_n} \left\{ \sum_j \lambda_j\bar{\mu}_j + \frac{\ln Z(\lambda_1,...,\lambda_n)}{N} \right\}\,
 \end{equation}
 with
 \begin{equation}
Z(\lambda_1,...,\lambda_n)=\int_{\Omega} e^{-N\sum_j \lambda_j\mu_j(\mathbf{X})} d\mathbf{X}\,.
\end{equation}

In the present case, defining
\begin{equation}
\label{eq:kappa}
\kappa=\frac{1}{N}\sum_{i=1}^{N}\left(1-\cos(p_i)\right)
\end{equation}
we can write Eqn.~\eqref{eq:ham} (apart from unessential constants) as
\begin{equation}
 \frac{H}{N}=\kappa-\frac{J}{2}(m_x^2+m_y^2)-\frac{K}{4}(m_x^4+2m_x^2m_y^2+m_y^4) 
\end{equation}
where $m_x$ and $m_y$ are the components of the magnetization vector 
$\mathbf{m}$ (see Eqn.~\eqref{eq:m}). The entropy of macrostates can be shown to 
be equal to:
\begin{equation}
\begin{aligned}
\bar{s}=\inf\limits_{\lambda_{\kappa}}& \lbrace\kappa\lambda_{\kappa} - \lambda_\kappa +\ln I_0(\lambda_\kappa)+\ln(4\pi^2)\rbrace + \\
&+ \inf\limits_{\lambda_x, \lambda_y} \lbrace\lambda_xm_x+\lambda_ym_y+\ln I_0\left(\sqrt{\lambda_x^2+\lambda_y^2}\right) \rbrace \\
=(\kappa&-1)B_{inv}(1-\kappa)+\ln I_0(B_{inv}(1-\kappa))+\\
&+\ln (4\pi^2)-mB_{inv}(m) +\ln I_0(B_{inv}(m))
\end{aligned}
\end{equation}
where $I_n(x)$ is the n-th Modified Bessel function of the first kind and 
$B_{inv}(x)$ is the inverse function of $I_1(x)/I_0(x)$. Let us notice that 
$\bar{s}$ depends actually only on $\kappa$ and $m=\sqrt{m_x^2+m_y^2}$. Now we 
can evaluate the microcanonical entropy $S(E)$ by using its relation with 
$\bar{s}(\kappa, m)$. The r.h.s. of Eqn.~\eqref{eq:entro} in the $N\gg 1$ limit 
can be estimated through a constrained extremal problem, namely as the supremum 
of $\bar{s}(\kappa, m)$ with the condition 
\begin{equation}
\label{eq:constr}
\bar{H}(\kappa, m_x, m_y)=E\,.
\end{equation}
It can be shown that such problem is fulfilled by the value $\tilde{m}$ of the 
magnetization that verifies the following condition:
\begin{equation}
\label{eq:solmicro-LD}
B_{inv}(\tilde{m})=(J\tilde{m}+K\tilde{m}^3)B_{inv}\left(1-\frac{E}{N}-\frac{J}{2}\tilde{m}^2-\frac{K}{4}\tilde{m}^4\right)\,.
\end{equation}
Once $\tilde{m}$ is known (Eqn.~\eqref{eq:solmicro-LD} can be solved by 
numerical methods) the corresponding value of $\kappa$ can be found from 
Eqn.~\eqref{eq:constr} and we finally have \begin{equation}
S(E)=N\bar{s}(\tilde{\kappa}, \tilde{m})\,.
\end{equation}
In a similar way we can derive the free energy of the system:
\begin{equation}
 F(\beta)=\inf_{\kappa,m}\left\{H(\kappa,m)-\frac{N}{\beta}\bar{s}(\kappa,m)\right\}\,.
\end{equation}

\section{Numerical integrator for the Langevin dynamics}
\label{sec:algo}

In this Appendix we briefly discuss the stochastic Verlet-like integration scheme 
 used to simulate Hamiltonian systems with generalized ``kinetic'' terms, 
subjected to a thermal bath at fixed $\beta$. We are therefore interested in 
integration schemes for the ``generalized'' Langevin equation
\begin{equation}
\label{eq:lang}
\mathbf{\dot{X}}=
 \begin{pmatrix}
  \dot{q}\\
  \dot{p}
 \end{pmatrix}
 =
  \begin{pmatrix}
  \partial_p H\\
  -\partial_q H-D\beta\partial_pH +\sqrt{2D}\xi(t)
 \end{pmatrix}
\end{equation}
 where $\xi(t)$ is a white noise with unitary 
variance and $D$ is a positive constant~\cite{baldovin18}.

Since in the limit $D=0$ the dynamics reduces to that of an isolated Hamiltonian 
system, we would like to find an algorithm that is symplectic in this limit. 
This problem has been addressed in \cite{melchionna07} for the case of quadratic kinetic 
energy: in the following we will apply the same reasoning to a wider class of 
Hamiltonian systems.

It can be shown \cite{honeycutt92} that the integration of Eqn.~\eqref{eq:lang} over a 
time-step $h$ leads to errors of order $h^{3/2}$ if the deterministic and the 
stochastic parts are evolved independently. In order to improve the stability of 
the algorithm one can alternate the integrations of positions and momenta, as in 
the usual Position Verlet. In the present case, the integration of the positions 
can be trivially written as
\begin{equation}
q(t_0+h)=q_0+\partial_p K(p_0) \frac{h}{2} + O(h^2)
\end{equation}
where $(q_0, p_0)$ is the state for the considered degree of 
freedom at time $t_0$ and $K(p)$ is the kinetic term.

On the other hand, in order to evolve the momenta we have to deal with the following partial differential equation:
\begin{equation}
\label{eq:pde}
 \dot{p}=F_0-D\beta\partial_pK(p) +\sqrt{2D}\xi(t)
\end{equation}
where $F_0=-\partial_q V(t_0)$). If $K(p)=p^2/2m$, such equation can be solved exactly; in the general case we can use the approximation
\begin{equation}
 \partial_pK(p)= \partial_pK\Bigr|_{p_0} + \partial_p^2 K\Bigr|_{p_0} (p-p_0)\,,
\end{equation}
with an error of order $h^2$ on $p(h)$. Substituting the above expression into Eqn.~\eqref{eq:pde} we get an equation of the form
\begin{equation}
 dp=(A + B p)dt +\sqrt{2D} dw\,,
\end{equation}
 where A and B are coefficients that are constant during the integration step. Solving the above equation, one gets the following Position Verlet-like integration scheme:
 \begin{equation}
 \begin{aligned}
  q^{\star}&=q_0+\partial_p K\Bigr|_{p_0} \frac{h}{2}\\
  p_h&=e^{Bh}p_0 + \frac{A}{B}(e^{Bh}-1) + \sqrt{2D} \mathcal{N}\left(\frac{1}{2B}(e^{2Bh}-1)\right)\\
  q_h&=q^{\star}+\partial_p K\Bigr|_{p_h} \frac{h}{2}
 \end{aligned}
\end{equation}
where
\begin{equation}
  \begin{aligned}
   A&= -\partial_q H\Bigr|_{q^{\star}} -D\beta\left(\partial_pK\Bigr|_{p_0} - \partial_p^2 K\Bigr|_{p_0} p_0\right)\\
   B&= - D\beta \partial_p^2 K\Bigr|_{p_0}\,.
  \end{aligned}
 \end{equation}
An equivalent Velocity Verlet-like algorithm can be found by inverting the integration order.
Let us notice that in the deterministic limit $D=0$ such second-order integrators are also symplectic.

\bibliography{biblio}

\end{document}